\documentclass[12pt,onecolumn]{IEEEtran}

\usepackage{graphicx,amsmath,amssymb,subfigure}

\newtheorem{theorem}{Theorem}

\begin{document}

%\begin{frontmatter}

%\title{Molecular Communication}
%\author{Sachin Kadloor}
\title{Molecular Communication Using Brownian Motion with Drift}

\author{Sachin Kadloor, Raviraj S. Adve, and Andrew W. Eckford%
\thanks{Manuscript received Jan. 12, 2010; revised Jun. 23, 2011; accepted Jul. 19, 2011. The associate editor coordinating the review of this paper and approving it for publication was Dr.~M.~Hughes. The material in this paper was presented in part at the  International Conference of Computer Communications and Networks (ICCCN), San Fransisco, CA, 2009.}%
\thanks{S. Kadloor was with The Edward S. Rogers Sr. Department of Electrical and Computer Engineering,
University of Toronto.  He is now with the Coordinated Science Laboratory, 
University of Illinois at Urbana-Champaign,
1308 West Main Street, Urbana, Illinois, USA 61801-2307.  Email: kadloor1@uiuc.edu}%
\thanks{R. S. Adve is with The Edward S. Rogers Sr. Department of Electrical and Computer Engineering,
University of Toronto, 10 King's College Road, Toronto, Ontario, Canada M5S 3G4.  Email: rsadve@comm.utoronto.ca}%
\thanks{A. W. Eckford is with the Department of Computer Science and Engineering, York University,
4700 Keele Street, Toronto, Ontario, Canada M3J 1P3.  Email: aeckford@yorku.ca}%
}

\maketitle

%\renewcommand{\baselinestretch}{1.5}
%\normalsize

\begin{abstract}
Inspired by biological communication systems,
molecular communication has been proposed as a viable scheme to
communicate between nano-sized devices separated by a very short
distance. Here, molecules are released by the transmitter into the
medium, which are then sensed by the receiver. This paper develops a
preliminary version of such a communication system focusing on the
release of either one or two molecules into a fluid medium with drift. We
analyze the mutual information between transmitter and the receiver
when information is encoded in the time of release of the molecule. 
Simplifying
assumptions are required in order to calculate the mutual information,
and theoretical results are provided to show that these calculations
are upper bounds on the true mutual information. Furthermore, optimized
degree distributions are provided, which suggest transmission
strategies for a variety of drift velocities.
\end{abstract}

%\end{frontmatter}

\section{Introduction}

Communications research has almost exclusively focused on
systems based on electromagnetic propagation. However, at scales
considered in nano-technology, it is not clear that these methods 
are viable.
Inspired by the chemical-exchange communication performed by
biological cells,
this paper considers {\em molecular communication}~\cite{hiy05},
in which information is
transmitted by an exchange of molecules. Specifically we consider the
propagation of individual molecules between closely spaced transmitters
and receivers, both immersed in a fluid medium. The transmitter encodes
a message in the pattern of release of the molecules into the medium;
these molecules then propagate to the receiver where they are sensed.
The receiver then attempts to recover the message by observing the
pattern of the received molecules.

It is well known that microorganisms exchange information by molecular
communication, with {\em quorum sensing}~\cite{bro01} as but one
example, where bacteria exchange chemical messages to estimate the
local population of their species. The biological literature on
molecular communication is vast, but there has been much recent work
concerning these systems as engineered forms of communication. Several
recent papers have described the design and implementation of
engineered molecular communication systems, using methods such as:
exchanging arbitrary molecules using Brownian motion in free
space~\cite{cav06}; exploiting gap junctions between cells to exchange
calcium ions~\cite{nak05,nak07}; and using microtubules and molecular
motors to actively drive molecules to their
destination~\cite{eno06,hiy08}. A comprehensive overview of the
molecular communication system is also given by~\cite{aky08,HiyamaMoritani} and the
references therein.

Given this engineering interest, it is useful to explore the
theoretical capabilities of molecular communication systems. To the
authors' knowledge, the earliest effort towards information-theoretic
analysis of these channels was given in~\cite{tho03}, which examined
information flow in continuous diffusion. In~\cite{eck07,eck-arxiv},
physical models and achievable bounds on information rate were provided
for diffusion-based systems. Information rates were provided
in~\cite{ata07,ata08} for the case where the receiver chemically
``reacts'' with the molecules and form ``complexes''.  In \cite{GarveyThesis},
it was shown that the additive white Gaussian noise (AWGN) is appropriate for
diffusion-based counting channels.
Information-theoretic results have also been obtained for specific
systems, such as propagation along microtubules~\cite{eck09, moo09b}
and continuous diffusion \cite{PierobonAkyildiz}.
All these studies indicate that useful information rates can be
obtained, although much lower per unit time than in electrical
communication; this is not surprising, since chemical processes are far
slower than electrical processes.
It is worth pointing out that these results build on theoretical work in Poisson and queue-timing channels
\cite{Kabanov,BitsThroughQueues}, which is an active area of research in information theory.

In any communication system, the potential rate of communication is
determined by the characteristics of the channel. We consider molecular
propagation in a fluid medium, governed by Brownian motion and,
potentially, a mean drift velocity. Our model is therefore applicable
to communications in, e.g., a blood vessel. This drift velocity is a
key difference between our work and \cite{eck07, eck-arxiv}, which
considered a purely diffusion channel. Furthermore, we
consider two cases - a single and two molecules being released. In this
regard, it is worth emphasizing the preliminary and conceptual nature
of this work. The long-term goal of this work is to understand the role
of both timing and the number of molecules (`amplitude'). Thus, the
contributions of this paper include:
\begin{itemize}
\item Calculation and optimization, under some simplifying
    assumptions, of mutual information in Brownian motion with
    drift, where the transmitter uses pulse-position modulation;
\item  Optimization of the degree distributions related to two transmit
    molecules; and
\item Demonstration (via theoretical results) that our simplified mutual
    information calculation is an {\em upper bound}	on the true
    mutual information of any practical implementation of this system.
\end{itemize}
Our optimized degree distributions reveal interesting features of these
channels, suggesting transmission strategies for system designers.

The paper is organized as follows. In Section~\ref{sec:sysmodel} we
describe the system under consideration, in which the propagation of
the molecule is analyzed and the probability distribution function of
the absorption time is derived. In Section~\ref{mutinf}, we
characterize the maximum information transfer per molecule, for the
case where information is encoded in the time of release of the
molecule, and the case of two molecules. In Section~\ref{sec:sim},
numerical and theoretical results arising from these models (including
optimized degree distributions) are presented. The paper wraps up with
some conclusions and suggestions for future work.

\section{System Model}\label{sec:sysmodel}

\subsection{Communication model} \label{CommModel}

\begin{figure}
\centering
	\includegraphics[width = 3.5in]{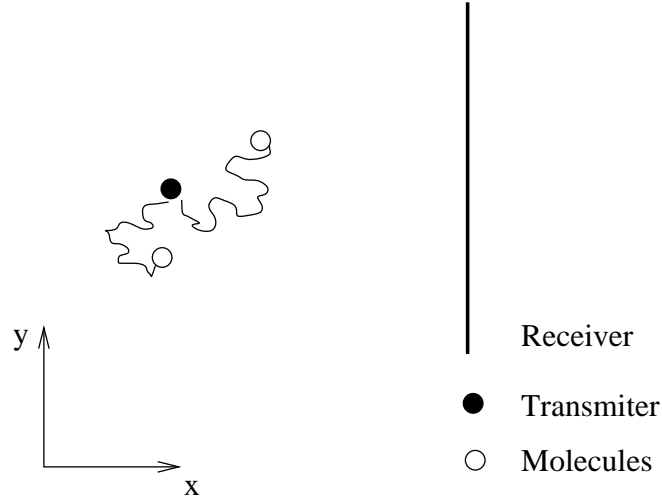}
	\caption{An abstract model of the molecular communication system. One or more molecules are released by the transmitter. These molecules then travel through the fluid medium to the receiver, which absorbs them upon reception. If all the molecules are identical, then information is conveyed from transmitter to the receiver only through the times at which the molecules are released.}
	\label{sysmodel}
\end{figure}

The communication model we consider is shown in Figure~\ref{sysmodel}.
The subsystems which make up the molecular communication system are:

\begin{enumerate}

\item {\bfseries Transmitter.} The transmitter is a source of
    identical molecules. It encodes a message in the time of
    dispersal of these molecules. We will assume that the
    transmitter can control precisely the time of dispersal of
    these molecules but does not influence the propagation of these
    molecules once dispersed.

\item {\bfseries Propagation medium.} The molecules propagate
    between transmitter and receiver in a fluid medium. The
    propagation is modeled as Brownian motion, and is characterized by two
    parameters: the drift velocity and the diffusion constant.
    These in turn depend on the physical properties of the fluid
    medium. The trajectories of different molecules are assumed to
    be independent of one another.

\item {\bfseries Receiver.} In this paper, the propagation of the
    molecule is assumed to be one dimensional. When it arrives at the receiver, the dispersed
    molecule is \emph{absorbed} by
    the receiver and \emph{is removed} from the medium. The
    receiver makes an accurate measurement of the time when it
    absorbs the molecule. It uses this information to determine the
    message sent by the transmitter.

\item {\bfseries Transmission of information.} The transmitter can
    encode information in either the time of dispersal of the
    molecules, or the number of molecules it disperses, or both.

\end{enumerate}

\begin{figure}[ht]
\centering
\subfigure[Modeling the motion of the particle as a one dimensional random walk]{
   \includegraphics[width=0.5\columnwidth] {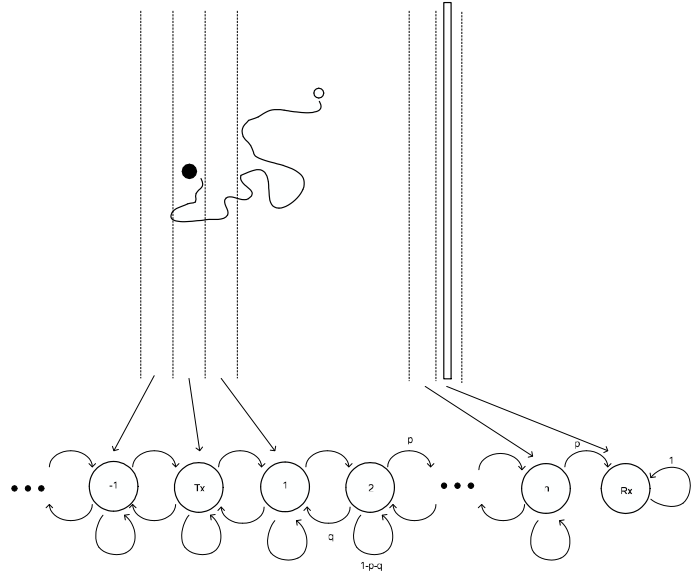}
   \label{fig:markovchain}
 }

 \subfigure[Sample paths of six particles in the same fluid medium, three released at $t=0$, three at $t=400$.]{
   \includegraphics[width=0.5\columnwidth] {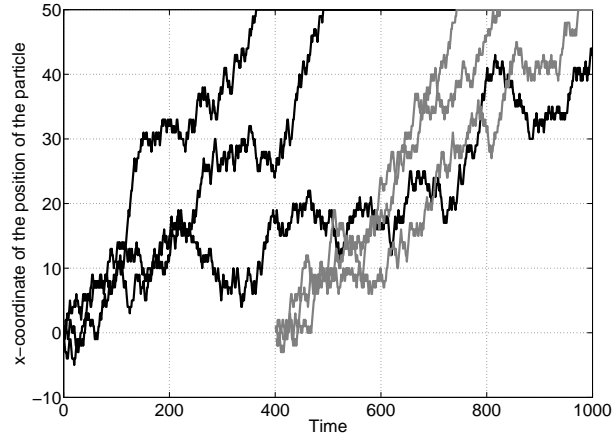}
   \label{fig:trajectories}
 }

\label{fig:sysmodel2}
\caption{If the size of the receiver is several orders greater than the size of the molecule, and if the velocity of the fluid in the `y axis' is negligible compared to the velocity of the fluid along `x axis' (Refer Fig. \ref{sysmodel}), then, one can ignore the y-coordinate of the position of the molecule and consider only the x-coordinate. The position of the molecule along the x axis is modeled as a Markov chain, specifically, as a one dimensional random walk. The bias of the walk (the values of $p$ and $q$) depend on the velocity of the fluid medium along the x-axis.}
\end{figure}

The motion of the dispersed molecule is affected by Brownian motion;
the diffusion process is therefore probabilistic and, in turn, the
propagation time to the receiver is random. Even in the absence of any
imperfection in the implementation of a molecular communication system,
this uncertainty in the propagation time limits the maximum information
rate per molecule. In this paper, we study the maximum information per
molecule that the transmitter can convey to the receiver, for a certain
velocity and diffusion in the fluid medium. Before proceeding to do so,
we need to characterize the propagation of the molecule in the medium.

\subsection{Diffusion via Brownian motion}\label{derivation}

Consider the discrete-time, discrete-space propagation model in
Figure~\ref{fig:markovchain}. Let $X(n)$ denote the position of the
particle at time $n$. Let $P_X(x,n;x_o,n_o)$ denote the probability
mass function (pmf) of the position of the particle at time $n$, given
that it was dispersed in the fluid medium at position $x_o$ at time
$n_o$. Assume that the fluid medium is static, and so the particle
disperses in either of the directions with equal probability. If $p$ is the probability that the particle moves from position $x$ to position $x+l$ in one time unit, and $q$ is the probability that it moves from position $x$ to $x-l$, then this situation is the case when 
$p=q=0.5$. It is easy to see that $P_X(x,n;x_o,n_o)$ obeys the equation
\begin{eqnarray}
 P_X(x,n+1;x_o,n_o) = \frac{1}{2}P_X(x-l,n;x_o,n_o) + \frac{1}{2}P_X(x+l,n;x_o,n_o), 	
	\label{eqn:balance}
\end{eqnarray}
which states that if a particle at time $n+1$ is at position $x$, then
at time $n$, it should have been at position $x-l$ or $x+l$, where $l$
is the distance between two slices of space. This formulation of
Brownian motion is analogous to a Wiener process, where distinct
increments of the motion are independent from each other.

Equation (\ref{eqn:balance}) can be re-written as
\begin{eqnarray}
	\nonumber
		\lefteqn{P_X(x,n+1;x_o,n_o)-P_X(x,n;x_o,n_o)} & & \\
	\nonumber
		& = & \frac{1}{2}(P_X(x-l,n;x_o,n_o)-P_X(x,n;x_o,n_o)) + \frac{1}{2}(P_X(x+l,n;x_o,n_o)-
                                               P_X(x,n;x_o,n_o)) \\
	                                       \label{eqn:diff_derivation_1}
		& = & \frac{l^2}{2}
		\left( \frac{1}{l} \left( \frac{P_X(x-l,n;x_o,n_o)-P_X(x,n;x_o,n_o)}{l} \right) \right) .
\end{eqnarray}
When $n\gg 1$ and $x \gg l$, the difference equation becomes a continuous time differential equation, yielding a probability distribution function (pdf) for the position of the particle, given by,
\begin{equation}
	\frac{\partial}{\partial n}P_X(x,n;x_o,n_o)
	= \frac{l^2}{2} \frac{\partial^2}{\partial x^2} P_X(x,n;x_o,n_o) .
	\label{eqn:diff_derivation}
\end{equation}
Now, considering a continuous time Brownian motion $X(t)$,
the probability density function of the position of the particle can be
modeled by the diffusion equation
\begin{equation}
	\frac{\partial }{\partial t}P_X(x,t;x_o,t_o) =
                    D \frac{\partial^2}{\partial x^2}P_X(x,t;x_o,t_o),
	\label{eqn:diff}
\end{equation}
where $D=l^2/2$ is the diffusion constant, whose value is dependent on
the viscosity of the fluid medium. Note that the above equation
characterizes only the `$x$-coordinate' of the position of the
molecule.  Solutions to this equation are well known.

%\subsection{Diffusion: Brownian motion with drift}\label{continuous}

Equation~(\ref{eqn:diff}) characterizes the motion $X(t)$ of the
particle in a macroscopically static medium. The more general and
useful case is that of a fluid medium is in motion with a mean drift
velocity $v$. Consider a frame of reference which is moving with the
same velocity. In this frame, the fluid medium is static and hence the
diffusion of the particle should obey Equation~(\ref{eqn:diff}). Let
$$x^{'}=x+vt, \quad t^{'}=t$$ be the new coordinate system, and without
loss of generality, assume $t_o=0$. Let
$$P_X(x,t;x_o,0)=P^{'}_{X^{'}}(x^{'},t^{'};x_o,0),$$ then
$$\frac{\partial }{\partial t^{'}}P^{'}_{X^{'}}(x^{'},t^{'};x_o,0) = D \frac{\partial^2}{\partial
x^{'2}}P^{'}_{X^{'}}(x^{'},t^{'};x_o,0).$$

In the static frame of reference, the differential equation can be written as
\begin{eqnarray}
\lefteqn{\frac{\partial }{\partial t}P^{'}_{X^{'}}(x^{'},t^{'};x_o,0)
        \frac{\partial t}{\partial t^{'}}+\frac{\partial }
        {\partial x}P^{'}_{X^{'}}(x^{'},t^{'};x_o,0)
        \frac{\partial x}{\partial t^{'}} =} \nonumber \\
        &&   D \frac{\partial}{\partial x^{'}}
            \left(\left(\frac{\partial x}{\partial x^{'}}
            \frac{\partial }{\partial x}+\frac{\partial t}{\partial x^{'}}
            \frac{\partial }{\partial t}\right)P^{'}_{X^{'}}(x^{'},t;x_o,0)\right) ,
\end{eqnarray}
which simplifies to
\begin{equation}
\frac{\partial }{\partial t}P_{X}(x,t;x_o,0) =
        \left(D \frac{\partial^2}{\partial x^{2}}+
                    v\frac{\partial }{\partial x} \right) P_{X}(x,t;x_o,0) .
	\label{eqn:diffwithdrift}
\end{equation}

\begin{figure}%
\centering
\includegraphics[scale=0.4]{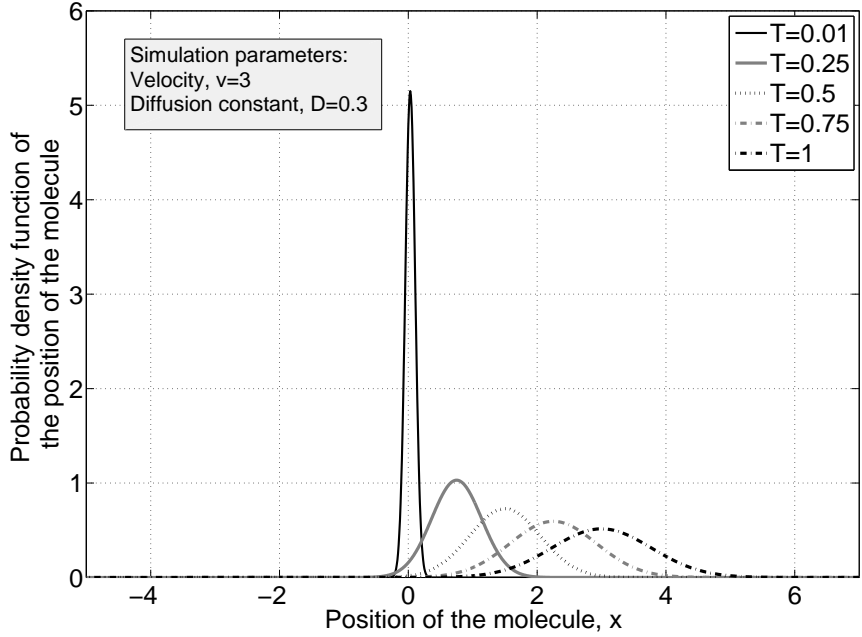}%
\caption{The pdf of the position of the molecule $P(x,t)$ for different values of $t$, when it is released at time $t=0$ at position $x=0$. Because of positive drift velocity, the mean of the pdf travels in the positive direction, and because of the diffusion, the variance of the pdf grows with time.}
\label{fig:pdf}
\end{figure}

Assume that there is no absorbing boundary (receiver) and that the fluid medium extends from $-\infty$ to $+\infty$. The probability density function of the location of the particle can be obtained by solving the differential Equation (\ref{eqn:diffwithdrift}) with boundary conditions $P_X(x,0;x_o,0)=\delta(x-x_o)$ and $P_X(\pm \infty,t;x_o,0)=0.$ The
solution to (\ref{eqn:diffwithdrift}) is given by~\cite{karatzas-book}
\begin{equation}
	P_X(x,t;0,0)
	= \frac{1}{\sqrt{4\pi Dt}} \mathrm{exp}\left(-\frac{(x-vt)^{2}}{4Dt}\right) .
	\label{eqn:gaussdiff}
\end{equation}
Equation (\ref{eqn:gaussdiff}) states that, for every $t$, the
probability density function (pdf) is a Gaussian centered at $vt$ with
variance $2Dt$. As expected, the expected location of the particle drifts along the direction of flow of the fluid medium with velocity $vt$. Figure \ref{fig:pdf} plots $P(x,t)$ for the case when $v=3$ and $D=0.3$. Furthermore, for any transmitter point $\zeta$ and transmit time $t_0$,
we have that
\begin{equation}
	P_X(x,t;\zeta,t_0)
	= \frac{1}{\sqrt{4\pi D (t-t_0)}} \mathrm{exp}\left(-\frac{((x-\zeta)-v(t-t_0))^{2}}
                                                                    {4D(t-t_0)}\right) .
	\label{eqn:gaussdiffint}
\end{equation}
As expected, Brownian motion $X(t)$ satisfying
(\ref{eqn:gaussdiff})-(\ref{eqn:gaussdiffint}) is a Wiener process with
drift.

 Now, consider the case when there is an absorbing surface (receiver) at $x=0$. The particle is absorbed and is removed from the system when it hits the absorbing surface. For such a system, to solve for $P_X(x,t;-\zeta,0))$, we need to solve the differential equation in (\ref{eqn:diffwithdrift}) with the following boundary conditions.
\begin{itemize}
	\item For $x<0, \quad P_X(x,0;-\zeta,0) = \delta(x+\zeta)$. The probability density function has a physical interpretation only for $x<0$. In this region, we require it to be a delta function at $t=0$ centered at $x=-\zeta$.
	\item $P_X(-\infty,t;-\zeta,0) = 0, \quad \forall t$.
	\item $P_X(0,t;-\zeta,0) = 0, \quad \forall t$. Condition imposed by the absorbing surface.
\end{itemize}

The solution to the differential equation can be computed using the \emph{method of images}, it is given by:
\begin{equation}
	P_X(x,t;-\zeta,0) = \frac{1}{\sqrt{4\pi Dt}} \mathrm{exp}\left(-\frac{(x+\zeta-vt)^{2}}{4Dt}\right)-  \frac{1}{\sqrt{4\pi Dt}} \mathrm{exp}\left(-\frac{(x-\zeta-vt)^{2}}{4Dt}\right) \mathrm{exp}\left(\frac{v\zeta}{D}\right)
	\label{eqn:eqnwithabs}
\end{equation}

\subsection{Distribution of Absorption Time}

Recall from Section \ref{sec:sysmodel} that the receiver senses the
particles only when they arrive, at which time they are absorbed and
removed from the system. Thus, for the purposes of this paper, the most
important feature of the Brownian motion $X(t)$ expressed in
(\ref{eqn:diffwithdrift})-(\ref{eqn:gaussdiffint}) is the {\em first
passage time} at the destination. For a Brownian motion $X(t)$, and an
absorbing boundary located at position $\zeta$, the first passage time
$\tau(\zeta)$ at the barrier is defined as
\begin{equation}
	\label{eqn:firstpassage}
	\tau(\zeta) = \min_{t} \{ X(t) \: : \: X(t) = \zeta \} .
\end{equation}
In Figure \ref{fig:trajectories}, the simulated trajectories of six particles, modeled as a random walk, through a medium are plotted. The particles were all released at $x=0$, three at time $0$ and three at time 400, into a fluid medium that had a positive drift velocity. The receiver is located at $x=50$. Notice the large variation in the absorption times. Among the particles released at $t=0$, one gets absorbed at $t\approx 360$, other at $t\approx 500$, and another does not get absorbed even by $t=1000$. Furthermore, this plot shows how particles can get absorbed in an order different from the order in which they were released. It is therefore important to understand the variation in the propagation times of the particle. 

The derivation of the first passage time for our case is given in \cite{chh-book}. Here, we repeat briefly the steps involved. At a given time $t$, the probability that the particle has not yet been absorbed is given by 
\begin{eqnarray}
	\bar{F}(t) &=& \int_{-\infty}^{0} P_X(x,t;-\zeta,0) dx \nonumber \\
	&=& \int_{-\infty}^{0} \frac{1}{\sqrt{4\pi Dt}} \mathrm{exp}\left(-\frac{(x+\zeta-vt)^{2}}{4Dt}\right) dx \nonumber \\ 
	&& -  \mathrm{exp}\left(\frac{v\zeta}{D}\right) \int_{-\infty}^{0} \frac{1}{\sqrt{4\pi Dt}} \mathrm{exp}\left(-\frac{(x-\zeta-vt)^{2}}{4Dt}\right) dx \nonumber \\
	&=& \left(1-\int_{\frac{-(vt-\zeta)}{\sqrt{2Dt}}}^{\infty} \frac{1}{\sqrt{2\pi}} e^{\frac{-x^2}{2}} dx \right) - \mathrm{exp}\left(\frac{v\zeta}{D}\right) \left(1- \int_{\frac{-(vt+\zeta)}{\sqrt{2Dt}}}^{\infty} \frac{1}{\sqrt{2\pi}} e^{\frac{-x^2}{2}} dx\right) \nonumber
\end{eqnarray} 
$\bar{F}$(t) is the probability that the particle has not been absorbed until time $t$. The probability that the particle has been absorbed before $t$ is given by $F(t)=1-\bar{F}(t)$. Hence, the probability density function of the absorption time is $f(t)=F^{'}(t)=-\bar{F}^{'}(t)$. 
\begin{eqnarray}
	f(t)&=& -\frac{d\bar{F}}{dt} \nonumber\\
	&=& \left( \frac{d}{dt}\int_{\frac{-(vt-\zeta)}{\sqrt{2Dt}}}^{\infty} \frac{1}{\sqrt{2\pi}} e^{\frac{-x^2}{2}} dx \right) - \mathrm{exp}\left(\frac{v\zeta}{D}\right) \left(\frac{d}{dt} \int_{\frac{-(vt+\zeta)}{\sqrt{2Dt}}}^{\infty} \frac{1}{\sqrt{2\pi}} e^{\frac{-x^2}{2}} dx\right) \nonumber \\
	&=& - \frac{1}{\sqrt{2\pi}} \mathrm{exp}\left(\frac{-(vt-\zeta)^2}{4Dt}\right)  \left( \frac{-v}{\sqrt{2Dt}} +  \frac{(vt-\zeta)}{2\sqrt{2Dt^3}} \right) + \nonumber\\
	&&\mathrm{exp}\left(\frac{v\zeta}{D}\right) \frac{1}{\sqrt{2\pi}} \mathrm{exp}\left(\frac{-(vt+\zeta)^2}{4Dt}\right)  \left( \frac{-v}{\sqrt{2Dt}} + \frac{(vt+\zeta)}{2\sqrt{2Dt^3}} \right) \nonumber\\
	&=& \frac{\zeta}{\sqrt{4\pi Dt^3}} \mathrm{exp}\left(\frac{-(vt-\zeta)^2}{4Dt}\right) 
	\label{eqn:abstime}
\end{eqnarray}

\begin{figure}%
\centering
\includegraphics[scale=0.4]{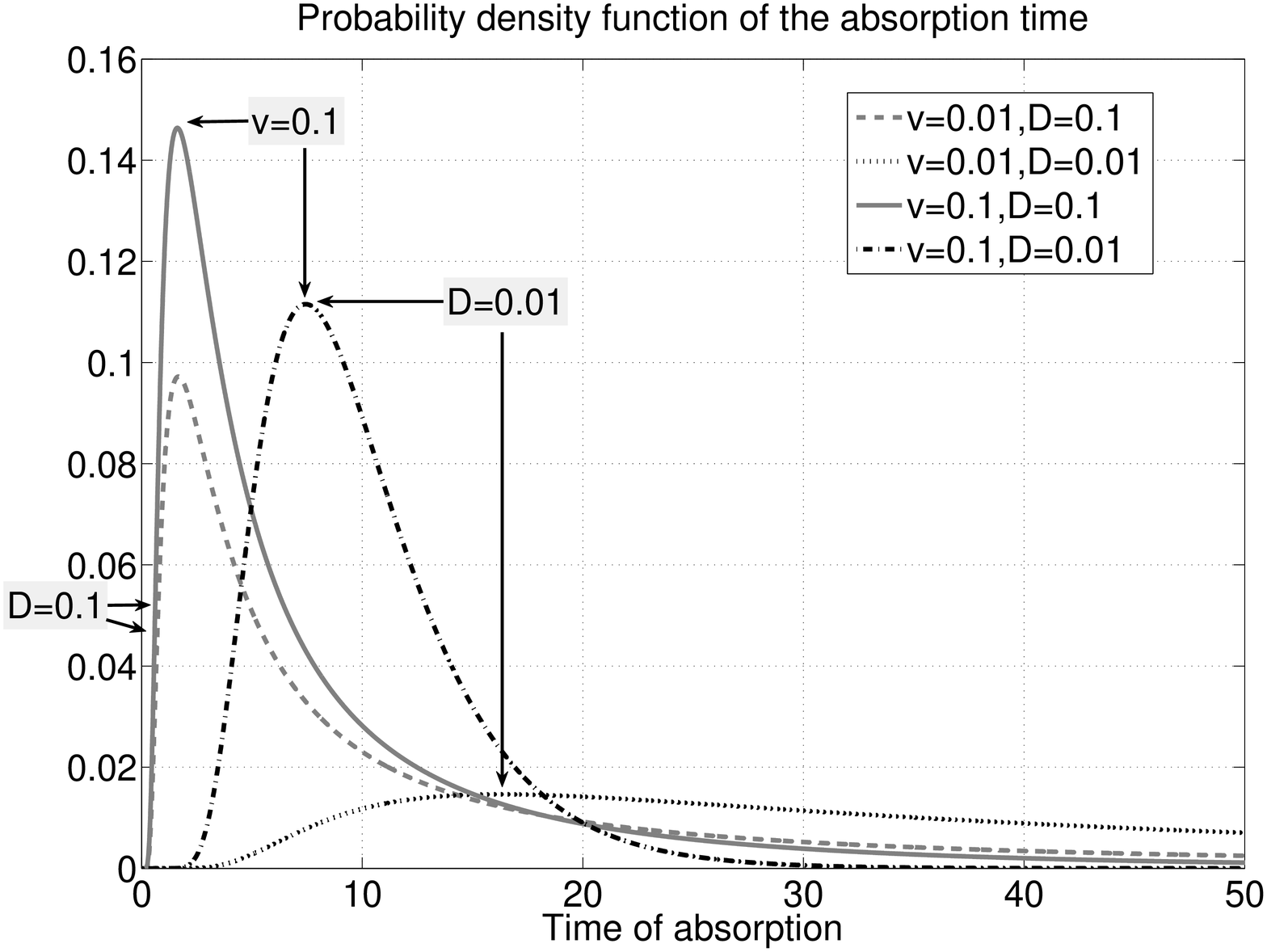}%
\caption{The time at which the molecule gets absorbed by the receiver, given that it was released at time 0, is a random variable. This is a result of the diffusion of the fluid medium. Here, we plot the probability distribution function of the absorption time for
different sets of velocity and diffusion. For this plot, the distance between the transmitter and the receiver is set at 1 unit.}
\label{fig:veldif}
\end{figure}

To summarize, (\ref{eqn:abstime}) gives the probability density function of the absorption time of a particle released in a fluid medium with diffusion constant $D$, at a distance $\zeta$ from the receiver, when the fluid has a constant velocity $v$. Note that this equation is valid only for positive drift velocities, i.e., when the receiver is downstream from the transmitter. Since our
communication is based largely on the time of transmission (and
reception), this pdf characterizes the \emph{uncertainty} in the
channel, and plays a role similar to that of the noise distribution
in an additive noise channel. Some example plots of this function are given in
Figure~\ref{fig:veldif}.

%\subsection{Communication in this media}\label{challenges}

\section{Mutual Information}\label{mutinf}

The transmitter encodes the message in the time of release of molecules
and possibly the number of molecules. Based on the number and the time
of absorption of the molecules, the receiver decodes the transmitted
information. This section develops the mutual information between the
transmitter and receiver for two cases: with a single transmitted
molecule and two molecules whose release times can be chosen
independently. For a given information transmission strategy at the transmitter
(called the {\em input distribution} in the information theoretic literature), 
the mutual information is also the maximum rate at which
information may be conveyed using that strategy. (Mutual information 
is related to but distinct from the {\em capacity}, which
is the maximum mutual information over all possible input distributions.)

\subsection{Overview}

In a traditional wireline communication system, 
receiver noise causes uncertainty in the reception, limiting the rate
at which information can be conveyed. 
%It may be a similar case with the
%system model considered above, where the receiver might record the
%number and time of absorption of the molecules incorrectly. 
However, as
discussed before, the uncertainty in the propagation time is a major
bottleneck to the information transfer in molecular communication. 
This uncertainty in the
propagation time also means that the order in which molecules are
received at the receiver need not be the order in which they were
transmitted. This will result in ``inter-block interference''. This is a
serious impairment in the low velocity regime, where the pdf of the
absorption time decays very slowly, making inter-block interference more likely.

%Achieving time synchronization between the transmitter and the receiver
%is not as straightforward as in the case of a wireline communication
%system. 
%In a wireline system, the transmitter sends a known signal to
%the receiver to mark the start of transmission. The same principle can
%obviously not be used as is. One way to synchronize the clocks is to do
%so before installing the transmitter and receiver. If the data rates
%are slow enough, the clocks need not be re-synchronized (to correct for
%difference in the oscillator frequencies) often.

In this paper, we ignore inter-block interference and assume that the
clocks are synchronized. Developing techniques for both issues are
significant works in themselves and outside the scope of this paper. So
our results are most relevant to system in a
fluid with some significant drift; further, as we show in Section \ref{sec:cap},
our results can be used to obtain upper bounds on both mutual information and capacity
for any drift velocity.

The channel here falls under a class of timing channels, channels where the mode of communication is through the timings of various events. The capacity of such channels are usually more difficult to characterize. A celebrated result in this field is the computation of the capacity of a single server queuing system \cite{BitsThroughQueues}. The molecular communication channel can be modeled as a $\cdot/G/\infty$ queuing system, i.e., an infinite server queuing system where the service time of a server is a random variable with distribution same as the pdf of the absorption time. To our knowledge, the exact capacity of such a channel has not been computed to date. 

\subsection{Single molecule: Pulse position modulation}

We first analyze the case of the transmitter releasing just a single
molecule. In such a scenario, it can encode information only in the
time of release of the molecule. The transmitter releases the molecule
(or not at all) in the beginning of one of $N$ time slots, each of unit
duration (i.e., $T_s = 1$ in arbitrary units); this
action on the part of the transmitter is called a {\em channel use}. 
This molecule then propagates through the
medium and is absorbed by the receiver in a later time slot. The
receiver then guesses the time slot in which the molecule was released.
This is a form of \emph{pulse-position modulation} (PPM).

Given that it has $(N+1)$ choices, the transmitter can encode a maximum
of $\textrm{log}_2 (N+1)$ bits of information per channel use, though in practice
much less due to the uncertain arrival times of the molecules. 
For instance, suppose the velocity of
the fluid medium is high enough so that the particle gets absorbed by
the receiver in $M \simeq N$ time slots with very high probability. 
In this case, one transmission strategy would be to emit a molecule in one of
$N/M$ time slots (each separated by $M$ slots), since inter-block interference
would thus occur with very low probability, and the transmitted information would
arrive without distortion.
For
such an ideal system, we can transmit information at a rate of
$\textrm{log}_2 N/M + 1$ bits per channel use. However, more practical
and interesting is the less than ideal case with lower velocities.

In this paper we neglect inter-block interference, i.e., we assume that
the receiver waits for enough time slots $M$, to ensure that the
molecule propagates to the receiver with high probability. Here $M$ is
chosen such that this probability is 0.999. Further, we assume that the
receiver sampling rate is $T_r = T_s/5$. This provides both a digital
input/output system while maintaining fairly high accuracy of the
received time. Both these parameters could be changed as required.

\subsection{Mutual information as an optimization problem}

Having dealt with preliminaries, we now derive the maximum possible
mutual information, here as an optimization problem. Define a random
variable $X$ to denote the time slot in which the transmitter releases
the molecule. Assume that the transmitter releases the particle at the
beginning of the $i^{th}$ slot ($1\leq i \leq N$) with probability
$p_i$. With probability $p_{0}=1-\sum_{i=1}^{N}p_i $, the transmitter does not
release the particle. Let $Y$ denote the time
slot in which the receiver absorbs the molecule. For the time being, we
allow $Y$ to range between $1$ and $\infty$, we will see shortly that
this is not required. Also, let $Y=0$ denote the event that the
molecule is never received. Since in our idealized case, the receiver
waits for a sufficiently long time, the event of $Y=0$ is the same
event that the molecule is not transmitted. Assume that the duration of
the time slot is $T_r$.

Let $F(t)$ denote the probability that the particle gets absorbed
before time $t$ given that it was released at time 0, i.e., $F(t)$ is
the integral of the pdf in \eqref{eqn:abstime}. Denote by $\alpha_j$
the probability that the particle arrives in the $j^{th}$ time slot,
given that it was released at time 0, which is equal to
$F(jT_r)-F((j-1)T_r)$ ; $\alpha_j = 0, j\leq 0 $. Let $H(X)$ denote the
entropy of random variable $X$ and let $\mathrm{entr}(x)$ represent
the binary entropy function, where
\begin{equation}
	\mathrm{entr}(x) = \begin{cases} -x\mathrm{log}_2 x & x>0 \\
	0 & x=0 .
	\end{cases}
\end{equation}
We now proceed to
calculate the mutual information between the random variables $X$ and
$Y$.
\begin{eqnarray}
	H(Y|X)&=& H(Y|X=0)p_0+\sum_{i=1}^{N} H(Y|X=i) p_i \nonumber\\
	&=& 0\times p_0+ \sum_{i=1}^{N} p_i \sum_{j=i+1}^{\infty} \mathrm{entr}\left(P(Y=j|X=i)\right)
	= \sum_{i=1}^{N} p_i \sum_{j=i+1}^{\infty} \mathrm{entr}\left(\alpha_{j-i}\right)\nonumber\\
	&=& (1-p_0)\sum_{k=1}^{\infty} \mathrm{entr}\left(\alpha_{k}\right), \quad \\
	H(Y)&=& \hspace*{-0.1in}\mathrm{entr}(P(Y=0))+\sum_{j=1}^{\infty}\mathrm{entr}(P(Y=j))\nonumber\\
	&=& \hspace*{-0.1in}\mathrm{entr}(p_0)+
                \sum_{j=1}^{\infty}\mathrm{entr}\left(\sum_{i=1}^{N}P(Y=j|X=i)p_i\right)\nonumber\\
	&=& \hspace*{-0.1in} \mathrm{entr}(p_0)+
            \sum_{j=1}^{\infty}\mathrm{entr}\left(\sum_{i=1}^{N}\left(\alpha_{j-i}\right)p_i\right)
\end{eqnarray}
\begin{eqnarray}
	I(X;Y) &=&H(Y)-H(Y|X) \nonumber \\
            &=& \hspace*{-0.1in}\mathrm{entr}(p_0)+
            \sum_{j=1}^{\infty}\mathrm{entr}\left(\sum_{i=1}^{N} p_i \alpha_{j-i}\right)
	-(1-p_0)\sum_{k=1}^{\infty} \mathrm{entr}\left(\alpha_{k}\right)
	\label{eqn:mut_inf}
\end{eqnarray}
As seen in Figure \ref{fig:veldif}, the sequence $\{\alpha_j\}$ is a an eventually decreasing sequence. The rate of decay
depends on the values of the drift velocity $v$ and the diffusion
coefficient $D$. The summations in (\ref{eqn:mut_inf}) can, therefore,
be terminated for some large enough $M$.

The expression for mutual information is a non-negative weighted sum of
concave functions plus a constant. Hence, the mutual information is a
concave function of the input distribution $\{p_i, i=1,\ldots,N\}$.
Finding the degree distributions, the values for $p_i$s which maximize
the entropy, is therefore a concave optimization problem. Standard
convex optimization techniques can therefore be used to solve for the
input probability distribution which maximizes the mutual information
efficiently, in particular, the Blahut-Arimoto algorithm \cite{Blahutarimoto,ArimotoBlahut}.

As a special case, suppose that we were to convey information only in
the time of release of the molecule, i.e., we require the molecule to
be transmitted. The derivation of mutual information is very similar to
the derivation above. Mutual information can then be expressed as
\begin{equation}\label{eqn:withoutp0}
I(X;Y)= \sum_{j=1}^{M}\mathrm{entr}\left(\sum_{i=1}^{N} p_i \alpha_{j-i}\right) -
                                        \sum_{j=1}^{M} \mathrm{entr}\left(\alpha_{j}\right),
\end{equation}
and, again, the optimal degree distribution can be obtained through
concave optimization.

\subsection{Two molecules}\label{multiplemolecules}

In the work so far we have considered only the propagation of a single
molecule and the focus was on PPM-based communication. We now take a
step toward involving amplitude wherein the transmitter can release two identical
molecules. The analysis is simplified by assuming that the propagation
paths of these two molecules are independent. The transmitter releases
each of these molecules in one of the $N$ time slots or chooses not to
release it. Based on the arrival times of these molecules at the
receiver, the receiver estimates their release times. However,
because of the nature of the diffusion medium, different molecules can
take different times to propagate to the receiver. Hence, the molecules
can be absorbed in a different order than in which they were released:
a key difference between this channel and traditional additive noise channels. As a result, the amount of information that can be conveyed through the medium with two indistinguishable molecules, as we will shortly see, is less than twice the amount of the information that can be conveyed using a single molecule. 

To obtain the maximum mutual information, let $X_1\in\{1,2,\ldots,N\}$
be the time slot in which the first particle is released,
$X_2\in\{X_1,X_1+1,\ldots,N\}$ be the time slot in which the second
particle is released. Let $Y_1$, and $Y_2$ be the time slots in which
the first and second particles are received. For notational
convenience, if a particle is not released, we denote it by a release
in slot 0. Likewise, if a particle is not received at the receiver, we
denote it by a reception in time slot 0.

The probability mass function of the reception times
$(P(Y_1,Y_2))$, and the conditional probability mass function
of the reception times given the transmission times
$(P(Y_1,Y_2|X_1,X_2))$ can be expressed in terms of the conditional
probability mass function of the reception time of one
molecule, given its transmission time
$(P(Y_1=y_1|X_1=x_1)=\alpha_{y_1-x_1})$. Let $p_{x_1x_2}$ represent
$P(X_1=x_1,X_2=x_2)$.

\begin{eqnarray}
P(Y_1=y_1,Y_2=0|X_1=x_1,X_2=0)&=&\alpha_{y_1-x_1}, \quad x_1,y_1>0\nonumber \\
P(Y_1=y,Y_2=y|X_1=x_1,X_2=x_2)&=&\alpha_{y-x_1}\alpha_{y-x_2}, \quad x_1\geq x_2,y>0 \nonumber\\
P(Y_1=y_1,Y_2=y_2|X_1=x_1,X_2=x_2)&=& \nonumber \\
&& \lefteqn{\alpha_{y_1-x_1}\alpha_{y_2-x_2}+ \alpha_{y_1-x_2}
                                            \alpha_{y_2-x_1}, \quad x_1 \geq x_2, y_1 \neq y_2>0} \nonumber\\
P(Y_1=0,Y_2=0)&=&p_{00}\nonumber \\
P(Y_1=y_1,Y_2=0)&=&\sum_{x_1=1}^{N}p_{x_10}\big( \alpha_{y_1-x_1} \big)\nonumber\\
P(Y_1=y,Y_2=y)&=&\sum_{x_1=1}^{N} \sum_{x_2=x_1}^{N} p_{x_1x_2}
\big(\alpha_{y-x_1}\alpha_{y-x_2}\big) \nonumber \\
P(Y_1=y_1,Y_2=y_2\ne y_1)&=&\sum_{x_1=1}^{N} \sum_{x_2=x_1}^{N} p_{x_1x_2} \big(\alpha_{y_1-x_1}
                                    \alpha_{y_2-x_2}+\alpha_{y_1-x_2}\alpha_{y_2-x_1}\big)\nonumber
\end{eqnarray}

The term $\alpha_{y_1-x_2}\alpha_{y_2-x_1}$ in the above equations
accounts for the event that the molecule released later gets absorbed
before the molecule which is released earlier. The mutual information
between the variables $(X_1,X_2)$ and $(Y_1,Y_2)$ can now be written in
terms of these probability mass functions. Note that in the
above derivation, we have assumed that $\alpha_{k}$ for $k\leq0$ is
defined as zero.

Using these equations, we can frame the mutual information maximization
as another optimization problem. The optimization is to be done over
the upper triangular $N\times N$ matrix $P_{X_1,X_2}(x_1,x_2)$, where
each entry in the matrix is a positive number and all the entries sum
to one. The mutual information is a concave function of the
optimization variables $\{ p_{x_1x_2}:
x_1\in\{1,2,\ldots,N\},x_2\in\{x_1,x_1+1,\ldots,N\} \}$. The exact
expression is tedious to write, and is omitted here.

\section{Results}
\label{sec:sim}

The well known Blahut-Arimoto algorithm \cite{Blahutarimoto,ArimotoBlahut} is used to compute, numerically, the input distribution that maximizes the mutual information in each of the different scenarios. The distance from the sender to the receiver, $\zeta$ is set to one unit in all the results presented here. 
\subsection{Release of a single molecule}
When one molecule is to be released, information can be conveyed in whether it is released or not, and if released, the slot number in which it is released. 

\begin{figure}%
\centering
\includegraphics[scale=0.4]{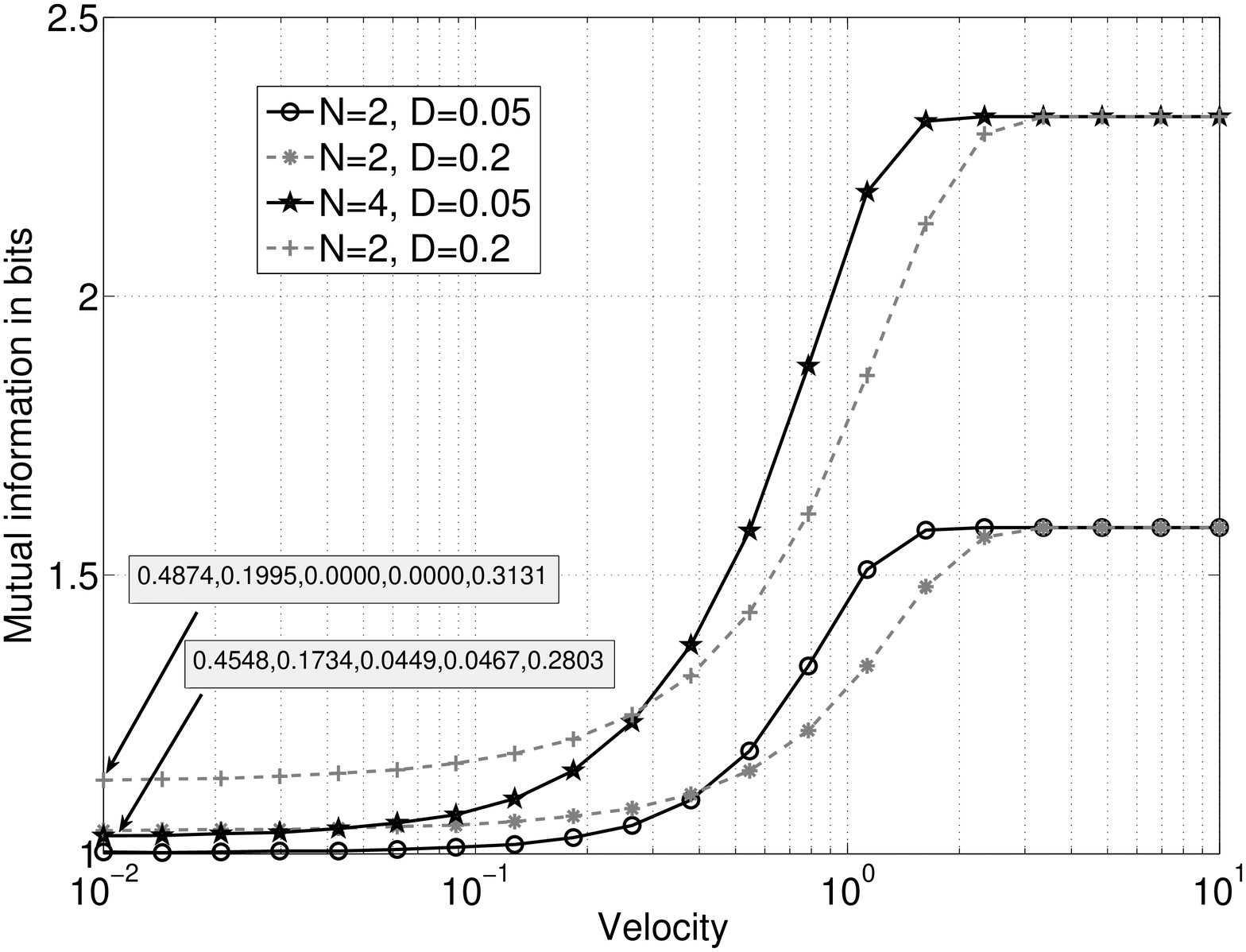}%
\caption{Variation of mutual information (which measures in bits, the information that can be conveyed from the transmitter to the receiver) with velocity. There are 2 sets of curves corresponding to the number of slots in which the molecule is released, $N=2$ and $N=4$. For $N=4$, we also list the p.m.f. of the release times which maximizes the mutual information.}%
\label{ratevsvel}%
\end{figure}

\paragraph*{Case when the molecule can be released in one of the $N$ slots, or not at all}
In Figure~\ref{ratevsvel}, we plot the mutual information as a function of velocity, for two
different sets of diffusion coefficients, 0.05, representing the low
diffusion scenario, and a high diffusion constant 0.2. We have two sets
of plots in the figure, one for the case where we have two slots in
which we can release the molecule, or choose not to release it, and
another, where we have four time slots. Also, we give the input
distribution $(p_0,p_1,p_2,p_3,p_4)$ at which the mutual information is
maximized at the two extreme values of velocity.

From the figure, it is evident that the mutual information increases
with an increase in velocity and saturates to a maximum of
$\mathrm{log}_2(N+1)$ bits. This trend is as expected. At high
velocities, the optimal, information maximizing, distribution is
uniform. This is because the receiver can detect, without error, the
slot in which the transmitter disperses the molecule. Also, because the
receiver waits for a sufficiently long time, we can detect, without
error, if a molecule was transmitted or not. Therefore, a lower limit
on the mutual information is one bit. At lower velocities, timing
information is completely lost and the mutual information is marginally
greater than one bit.

\begin{figure}%
\centering
\includegraphics[scale=0.4]{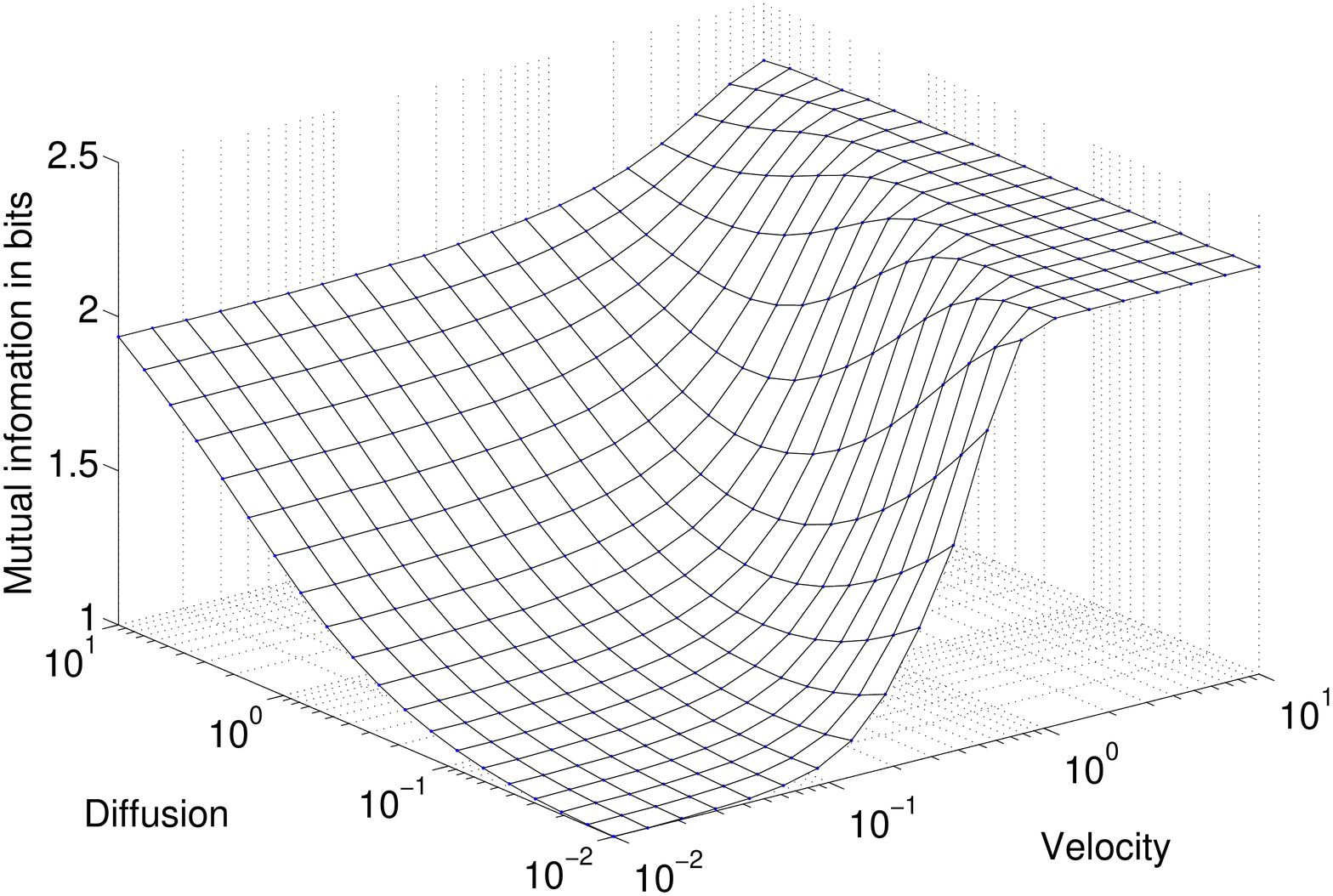}%
\caption{A grid plot denoting the mutual information for a range of different velocities and diffusion constants, for the case when $N=4$. Observe that at lower velocities, more information can be transferred in a medium with higher diffusion constant.}
\label{fig:grid}%
\end{figure}

The diffusion constant is a measure of the uncertainty in the
propagation time. Hence, we would expect the mutual information to be
lower when the diffusion constant is high. This is indeed the case at
high velocities. However, it is surprising that a higher diffusion
constant results in higher mutual information at low velocities (Also refer Figure \ref{fig:grid}). This
is because, at low velocities, it is the diffusion in the medium which
aids the propagation of the molecule from the transmitter towards the
receiver. This is illustrated in the pdf of the absorption time, shown
in Figure \ref{fig:veldif}. Compared to the case when the diffusion in
the medium is low, the probability distribution function is more
``concentrated'' (lower uncertainty) when the diffusion in the medium
is higher. Unfortunately, there does not seem to be a single parameter
that characterizes the resulting interplay between velocity and
diffusion.

\begin{figure}%
\centering
\includegraphics[scale=0.4]{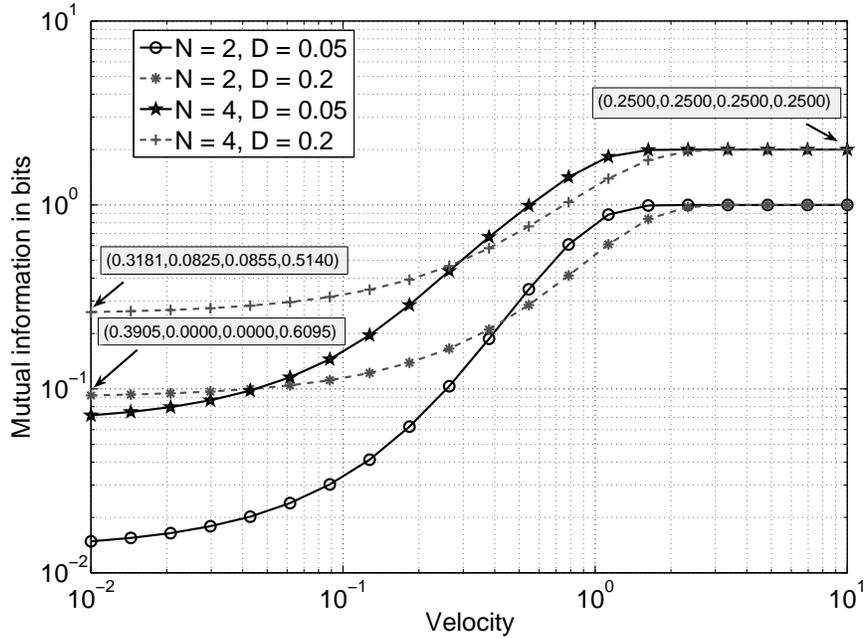}%
\caption{Variation of mutual information with velocity when the transmitter
must disperse the molecule ($p_0 = 0$). The scenario is similar to the one used in plotting Figure \ref{ratevsvel}, with the difference being that the transmitter is not permitted not to transmit a molecule. }
\label{withoutp0}%
\end{figure}

\begin{figure}%
\centering
\includegraphics[scale=0.4]{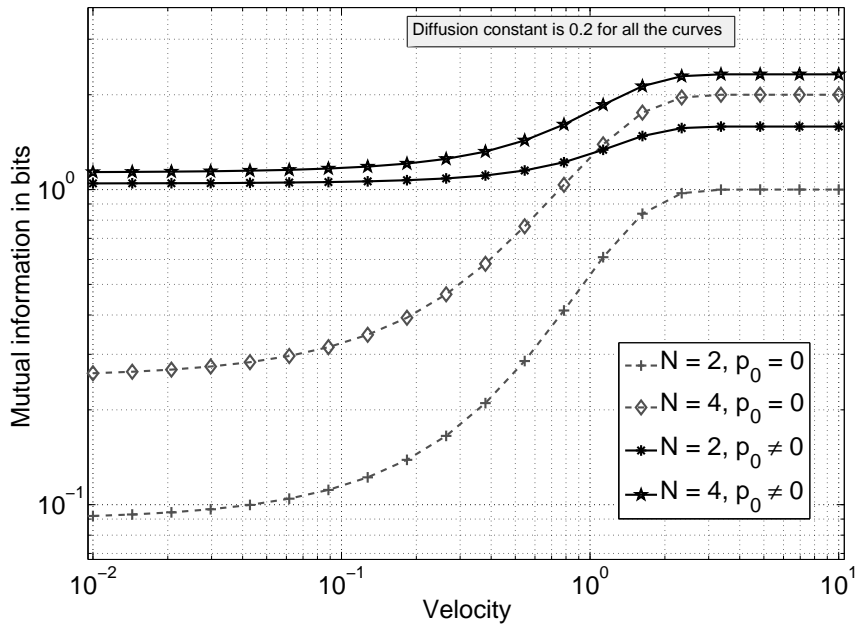}%
\caption{A comparison of the information bits conveyed in the scenarios when the transmitter must ($p_0 = 0$) or may not release the molecule. Plots from Figures \ref{ratevsvel} and \ref{withoutp0} are compared here.}%
\label{comp}%
\end{figure}

\paragraph*{Case when we do not permit the transmitter to not transmit the molecule}
The information, in this scenario, is conveyed only in the time of release of the molecule. We find the input distribution which maximizes
(\ref{eqn:withoutp0}). The mutual information in this case is plotted
in Figure \ref{withoutp0}. The maximum mutual information is now
$\mathrm{log}_2(N)$ bits, which is achieved at high velocities.
However, it is in the low velocity regime where the mutual information
is significantly lower than the case where the transmitter is allowed
to not transmit the molecule. Figure \ref{comp} compares the two
scenarios.

From the results, we see that the velocity-diffusion region can be
roughly classified into three regimes:
\begin{itemize}
\item A diffusion dominated region, where mutual information is
    relatively insensitive to the velocity; this corresponds to
    $v<10^{-1}$ in Figure~\ref{ratevsvel}.
\item A high-velocity region where the mutual information is
    insensitive to the diffusion constant; this corresponds to $v
    > 3$ in Figure~\ref{ratevsvel}.
	\item An intermediate regime, where the mutual information is
highly sensitive to the velocity and diffusion constant of the
medium, $10^{-1}<v<3$ in Figure \ref{ratevsvel}.
\end{itemize}

In the low velocity regime, we see no significant improvement in the
mutual information when we increase the number of time slots in which
we can release the molecules. As expected, very little information can
be conveyed in the time of release of the molecule when there is high
uncertainty in the propagation time. Hence, we need to explore
alternative ways of encoding message in this regime.

\subsection{Release of multiple molecules}
\begin{figure}%
\centering
\includegraphics[scale=0.6]{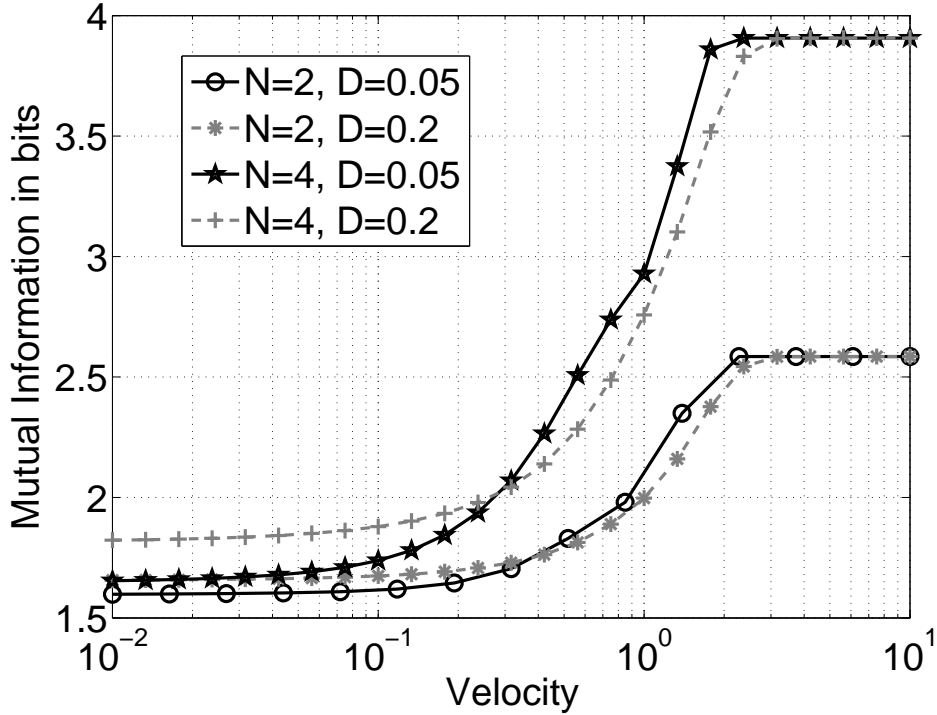}%
\caption{Variation of mutual information with velocity for the case when the transmitter is allowed to release at most two molecules. The mutual information maximizing input distributions at the extreme points of the graph are given in Tables \ref{t2low}, \ref{t2high}, \ref{t2high_upp}, \ref{t4low}, \ref{t4high} and \ref{t4high_upp}. }%
\label{two}%
\end{figure}

Here, we present the results of the scenario in which the transmitter is allowed to transmit at most 2 molecules. The results are presented in Figure \ref{two}. We have two sets of plots, one where the transmitter can
release the molecule in one of two time slots, other, where the
transmitter can release the molecule in one of four time slots. At low
velocities, the mutual information is close to $\mathrm{log}_2 3$ bits.
This is because, at low velocities, any information encoded in the time
of release of the molecule is lost. The receiver can however accurately
estimate the number of molecules transmitted. With two molecules, the
receiver can decode if the number of molecules transmitted was one or
two or zero. However, this is because, we wait for infinite time at the
receiver. The probability distribution function which attains the
maximum mutual information at low velocities assigns, roughly, a
probability of $\frac{1}{3}$ to the events of releasing one or two or
no molecules.

At very high velocities, information encoded in both the time and
number of molecules released is retained through the propagation.
Hence, a maximum of $\mathrm{log}_2 \frac{(N+1)(N+2)}{2}$ bits can be
conveyed at high velocities.

In Tables \ref{t2low}, \ref{t2high} and \ref{t2high_upp} we list the
mutual information maximizing input distributions for the case of
release of two molecules in \emph{two} time slots. Tables \ref{t4low},
\ref{t4high} and \ref{t4high_upp} list the input distributions for the
case of release of two molecules in \emph{four} time slots. As
expected, at low velocities, the total probability of releasing one,
two or zero molecules is roughly one third each. The molecules, to
minimize uncertainty, are transmitted `far apart'.

It is however surprising to note that for reasonable velocities when
two molecules are released, they are both to be released in the same
time slot. This may be explained by the fact that, due to diffusion,
molecules can arrive out of order and the timing information is lost.
Transmitting both molecules at once avoids this confusion. This is also
an important result; if this trend is to hold true for the release of
multiple molecules, then we could consider only those schemes wherein
all the molecules are released in one of the time slots, and where
information is encoded only in the time slot in which all the molecules
are released.

\begin{table}
\centering
\caption{Mutual information maximizing input distribution when two molecules are released in one of
the two possible slots or not released at all, $v=10^{-2}$, $d=0.05$}
\label{t2low}
\begin{tabular}{|c|c|c|c|}
\hline
         & $P(X_2=1)$ & $P(X_2=2)$ & $P(X_2=0)$ \\
\hline
$P(X_1=1)$ & 0.1424   & 0   & 0.1412	\\
$P(X_1=2)$ & 0   & 0.1939   & 0.1921	\\
$P(X_1=0)$ & 0   & 0   & 0.3303 \\
\hline
\end{tabular}

\caption{$v=10^{-2}$, $d=0.2$}
\label{t2high}
\begin{tabular}{|c|c|c|c|}
\hline
         & $P(X_2=1)$ & $P(X_2=2)$ & $P(X_2=0)$ \\
\hline
$P(X_1=1)$ & 0.1382   & 0   & 0.1299	\\
$P(X_1=2)$ & 0   & 0.2113   & 0.2035	\\
$P(X_1=0)$ & 0   & 0   & 0.3171 \\
\hline
\end{tabular}

\caption{$v=10$, $d=0.2$ or $0.05$}
\label{t2high_upp}
\begin{tabular}{|c|c|c|c|}
\hline
         & $P(X_2=1)$ & $P(X_2=2)$ & $P(X_2=0)$ \\
\hline
$P(X_1=1)$ & 0.1667   & 0.1667   & 0.1667	\\
$P(X_1=2)$ & 0   & 0.1667   & 0.1667	\\
$P(X_1=0)$ & 0   & 0   & 0.1667 \\
\hline
\end{tabular}

\end{table}

\begin{table}
\centering
\caption{Mutual information maximizing input distribution when two molecules are
released in one of the four possible slots or not released at all, $v=10^{-2}$, $d=0.05$}
\label{t4low}
\begin{tabular}{|c|c|c|c|c|c|}
\hline
         & $P(X_2=1)$ & $P(X_2=2)$ & $P(X_2=3)$ &$P(X_2=4)$ & $P(X_2=0)$ \\
\hline
$P(X_1=1)$ & 0.1395   & 0     & 0     & 0    & 0.1313	\\
$P(X_1=2)$ & 0   & 0     & 0     & 0    & 0  \\
$P(X_1=3)$ & 0   & 0     & 0     & 0    & 0  \\
$P(X_1=4)$ & 0   & 0     & 0     & 0.2094    & 0.2022  \\
$P(X_1=0)$ & 0   & 0     & 0     & 0    & 0.3176  \\
\hline
\end{tabular}

\caption{$v=10^{-2}$, $d=0.2$}
\label{t4high}
\begin{tabular}{|c|c|c|c|c|c|}
\hline
         & $P(X_2=1)$ & $P(X_2=2)$ & $P(X_2=3)$ &$P(X_2=4)$ & $P(X_2=0)$ \\
\hline
$P(X_1=1)$ & 0.1345   & 0     & 0     & 0    & 0.1256	\\
$P(X_1=2)$ & 0   & 0.0305     & 0     & 0    & 0.0122  \\
$P(X_1=3)$ & 0   & 0     & 0.0129     & 0   & 0  \\
$P(X_1=4)$ & 0   & 0     & 0     & 0.2052    & 0.1964  \\
$P(X_1=0)$ & 0   & 0     & 0     & 0    & 0.2827  \\
\hline
\end{tabular}

\caption{$v=10$, $d=0.2$ or $0.05$}
\label{t4high_upp}
\begin{tabular}{|c|c|c|c|c|c|}
\hline
         & $P(X_2=1)$ & $P(X_2=2)$ & $P(X_2=3)$ &$P(X_2=4)$ & $P(X_2=0)$ \\
\hline
$P(X_1=1)$ & 0.0667   & 0.0667     & 0.0667     & 0.0667    & 0.0667	\\
$P(X_1=2)$ & 0   & 0.0667     & 0.0667     & 0.0667    & 0.0667  \\
$P(X_1=3)$ & 0   & 0     & 0.0667     & 0.0667    & 0.0667  \\
$P(X_1=4)$ & 0   & 0     & 0     & 0.0667    & 0.0667  \\
$P(X_1=0)$ & 0   & 0     & 0     & 0    & 0.0667  \\
\hline
\end{tabular}

\end{table}

\section{Relationship to achievable information rates and capacity}
\label{sec:cap}

When pulse-position modulation is used, symbols are normally
transmitted consecutively.  That is, if the duration of a symbol is
$T$, then the first symbol is transmitted on the interval $[0,T)$, the
second on the interval $[T,2T)$, and so on. However, for the Brownian
motions considered in Section \ref{sec:sysmodel}, molecules transmitted
during a given interval may arrive during a later interval, causing
inter-block interference. In this paper, we have avoided this problem
by only considering symbols transmitted in isolation, disregarding inter-block interference.

In fact, for a fixed input distribution,
our information results lead to an {\em upper bound} on the
mutual information under consecutive symbol transmission. To show this,
let $\mathcal{X}$ represent the alphabet of allowed symbols. For
simplicity, suppose that a symbol is composed of the release of a
single molecule, although this assumption can be relaxed without
changing the argument.  Then we will assume that $\mathcal{X}$ is a
finite, discrete list of allowed molecule release times on the interval $[0,T)$;
the cardinality $|\mathcal{X}|$ gives the number of allowed release times.
Further, there exists a discrete input distribution, with pmf $p_X(x)$, over $\mathcal{X}$.
Let $\mathcal{Y}$ represent the
corresponding set of channel outputs, given a {\em single} symbol input
to the channel, and disregarding inter-block interference. 
Since $\mathcal{Y}$ is the arrival time of a single
molecule transmitted on the interval $[0,T)$, then clearly $\mathcal{Y}= [0,\infty)$, and nothing
changes if $\mathcal{Y}$ is quantized.

Let $\mathbf{x} = [x_1, x_2, \ldots, x_n] \in \mathcal{X}^n$ and
$\mathbf{y} = [y_1, y_2, \ldots, y_n] \in \mathcal{Y}^n$ represent
vectors of channel inputs and outputs, respectively, for $n$ uses of
the channel in isolation; throughout this section, we will assume that $x_i$ is independent and
identically distributed (IID) for each $i$. 
Suppose the symbols in $\mathbf{x}$ are
transmitted consecutively. Then the resulting sequence of molecule
release times can be written $\mathbf{r} = [r_1,r_2,\ldots]$, where
\begin{equation}
	\label{eqn:r}
	r_i = x_i + (i-1)T .
\end{equation}
Since $x_i \in [0,T)$, clearly $r_i \in [(i-1)T,iT)$.
Note that the mutual information per unit time of the
channel is given by $I(X;Y)/T$, which is calculated for given $\mathcal{X}$ and $p_X(x)$.

Let $u_i$ represent the arrival corresponding to $r_i$.  Since $r_i$ is
a time-delayed version of $x_i$, and $y_i$ is the arrival corresponding
to $x_i$, from (\ref{eqn:r}) we have
\begin{equation}
	\label{eqn:u}
	u_i = y_i + (i-1)T .
\end{equation}
The corresponding vector is $\mathbf{u} =
[u_1,u_2,\ldots,u_n]$.
However, the receiver does not observe $\mathbf{u}$ directly -- instead, it observes
$\mathbf{w}$, where
\begin{equation}
	\label{eqn:sort}
	\mathbf{w} = \mathrm{sort}(\mathbf{u}) ,
\end{equation}
and where the function $\mathrm{sort}(\cdot)$ sorts the argument vector in increasing order.
That is, while information symbol $x_i$ corresponds to arrival time $u_i$, it is potentially unclear
which element of $\mathbf{x}$ corresponds to arrival time $w_i$.

Since the length-$n$ vectors of consecutive input symbols 
$\mathbf{r}$ and sorted outputs $\mathbf{w}$ are random variables, we
can write the mutual information between them as
$I(\mathbf{R};\mathbf{W})$.  However, we are more interested in 
the mutual information per unit time $I(R;W)$, which is given by
\begin{eqnarray}
	\nonumber
	I(R;W) & = & \lim_{n \rightarrow \infty} \frac{1}{nT + \delta} I(\mathbf{R};\mathbf{W}) \\
	\label{eqn:truemutualinfo}
	& = & \lim_{n \rightarrow \infty} \frac{1}{nT} I(\mathbf{R};\mathbf{W}) - \epsilon ,
\end{eqnarray}
where $nT$ represents the total time to transmit $n$ symbols, $\delta$ is the extra time after $nT$
required to wait for all remaining molecules to arrive, and
\begin{equation}
	\label{eqn:epsilon}
	\epsilon = \left( \frac{1}{nT} - \frac{1}{nT + \delta} \right) I(\mathbf{R};\mathbf{W}).
\end{equation}
We let $\delta = \log n$, so that $\lim_{n \rightarrow \infty} \delta = \infty$
(which is long enough time for all molecules to arrive with probability 1).
%and $\lim_{n \rightarrow \infty} \epsilon = 0$ (which can be seen from (\ref{eqn:epsilon})).
%Then (\ref{eqn:truemutualinfo}) becomes
%%
%\begin{equation}
%	I(R;W) = \lim_{n \rightarrow \infty} \frac{1}{nT} I(\mathbf{R};\mathbf{W}) .
%\end{equation}
%%
With this in mind, we have the following result:
\begin{theorem}
	\label{thm:main}
	\begin{equation}
		\frac{1}{T}I(X;Y) \geq I(R;W) .
	\end{equation}
\end{theorem}
\begin{proof}
	From (\ref{eqn:truemutualinfo}), since $\epsilon$ is positive,
	\begin{displaymath}
		I(R;W) \leq \lim_{n \rightarrow \infty} \frac{1}{nT + \delta} I(\mathbf{R};\mathbf{W}) .
	\end{displaymath}
	Then we have
	that
	\begin{eqnarray}
		I(\mathbf{X};\mathbf{Y}) & = & I(\mathbf{R};\mathbf{U}) \\
		& \geq & I(\mathbf{R};\mathbf{W}) ,
	\end{eqnarray}
where the first equality follows from (\ref{eqn:r})-(\ref{eqn:u}),
since $\mathbf{r}$ and $\mathbf{u}$ are bijective functions of
$\mathbf{x}$ and $\mathbf{y}$, respectively; and the second inequality
follows from the data processing inequality (e.g., see \cite{cov-book})
and (\ref{eqn:sort}). Finally, since $x_i,y_i$ and $x_j,y_j$ are
independent for any $i \neq j$,	$I(\mathbf{X};\mathbf{Y}) = nI(X;Y)$,
and the theorem follows.
\end{proof}

Note that the result in Theorem \ref{thm:main} bounds the mutual information, and thus applies to {\em each}
set $\mathcal{X}$ and
input pmf $p_X(x)$; however, we can also
show that the result applied to capacity. Let $C_m$ represent the maximum of $I(X;Y)$ where $|\mathcal{X}| = m$,
i.e.,
\begin{equation}
	C_m = \max_{p_X(x) : |\mathcal{X}| = m} I(X;Y).
\end{equation}
The capacity of the channel uses in isolation is then given by
\begin{equation}
	C = \lim_{m \rightarrow \infty} C_m .
\end{equation}
It remains to show that this limit exists, which we do in the following result.
\begin{theorem}
	\label{thm:limit}
	$C$ exists, and is finite, if $0 < v,D,\zeta,T < \infty$.
	Furthermore, if $\max_{p_X(x)} I(R;W)$ represents the capacity of $I(R;W)$ under 
	IID inputs, then
	\begin{equation}
		\max_{p_X(x)} I(R;W) \leq \frac{C}{T} .
	\end{equation}
\end{theorem}
\begin{proof}
	To prove the first part of the theorem, we proceed in two steps.
	\begin{enumerate}
		\item {\em $C_m$ is a nondecreasing sequence.}
		For each $m$, either: the 
		maximizing distribution $p_X(x)$ 
		(or every maximizing distribution, if not unique)
		satisfies $p_X(x) > 0$ for all $x \in \mathcal{X}$;
		or $p_X(x) = 0$ for at least one $x \in \mathcal{X}$ (in at least one maximizing distribution, if not unique). 
		If the former is true, then $C_m > C_j$ for all $j < m$; if the
		latter is true, then $C_m = C_{m-1}$. Thus, $C_m$ is nondecreasing in $m$.
		
		\item {\em $C_m$ is upper bounded.}
		We can write $I(X;Y) = H(Y) - H(Y|X) = H(Y) - H(N)$, where $H(N)$ is the entropy of the first arrival time.
		We can upper bound $H(Y)$ independently of $m$ as follows. If the pdf of $y$ is $f_Y(y)$, then 
		$H(Y) = E[\log_2 1/f_Y(y)]$, where $E[\cdot]$ is expectation. If $g(y)$ is any valid pdf of $y$, then by a well-known
		property of entropy, $H(Y) \leq E[\log_2 1/g(y)]$ (with equality when $g(y) = f_Y(y)$). 
		Pick $g(y) = e^{-y}$ (supported on $y=[0,\infty)$), the exponential distribution with unit mean.
		Then $H(Y) \leq E[y] \log_2 e$, which is finite if $E[y]$ is finite. 
		Finally, $E[y] = E[x] + E[n] \leq T + E[n]$, and $E[n]$ is known to be finite
		if $0 < v,D,\zeta < \infty$ \cite{chh-book}.
	\end{enumerate}
	Since $C_m$ is a nondecreasing, upper bounded sequence, it must have a finite limit.
	
	To prove the second part of the theorem, note that Theorem \ref{thm:main} applies to all 
	input distributions $p_X(x)$; thus, it also applies to the one maximizing $I(R;W)$. 
	As a result, since $C$ exists and is finite (from the first part of the theorem), it
	is a nontrivial upper bound on $\max_{p_X(x)} I(R;W)$.
\end{proof}
In \cite{eck-arxiv}, it was shown that the mutual information cannot be tractably computed
in general for ``sorting'' channels, i.e., those with outputs given by (\ref{eqn:sort}).
Since $I(X;Y)$ can be calculated relatively easily, the results from Theorems \ref{thm:main}-\ref{thm:limit}
give us useful information about the capacity of a practical system.

\section{Conclusions and Future Work}\label{sec:futwork}

In this paper, a framework was constructed to study data rates that can
be achieved in a molecular communication system. A simple model was
considered for the communication system, consisting of a transmitter
and receiver separated in space, immersed in a fluid medium. The rates
achieved by a simple pulse-position modulation protocol were analyzed,
where information is encoded in the time of release of the molecule.
These results were extended to two molecules wherein the optimal
distribution reverted to the PPM protocol. While preliminary, the
results of this work suggest practical data transmission strategies
depending on the value of the drift velocity.

Given the preliminary nature of this work, there are many interesting
related problems. For example, it would be useful to consider the limitations 
inherent in molecular production and detection: precise control over release times
and amounts, and precise measurement of arrival times, may not be possible;
more realistic communication models could be produced.  Furthermore,
the communication architecture of molecular communication systems may be considered;
for instance, in order to achieve the mutual information results given in this paper, 
error-correcting codes must be
used; an appropriate modulation and 
coding strategy for molecular communication needs to be identified.  Finally, channel
estimation techniques need to be derived in order to cope with unknown parameters, such
as an unknown drift velocity.  Much work remains to be done to understand molecular
communication from a theoretical perspective, which presents many interesting and exciting
challenges to communication researchers.

\renewcommand{\baselinestretch}{1}
\normalsize

\bibliographystyle{IEEEtran}
\bibliography{particle}

\end{document}